# Effects of Flavor Dependence on Weak Decays of $J/\psi$ and $\Upsilon$

Rohit Dhir[¥], R.C. Verma and Avinash Sharma*

[¥]Department of Physics, Yonsei University, Seoul, South Korea;

*Department of Physics*, Punjabi University, Patiala -147001, India;

* *School of Basic and Applied Sciences, GGSIP University,*

*New Delhi-110075, India.*

e-mail: dhir.rohit@gmail.com

We carry out a detailed analysis of effects of flavor dependence of average transverse quark momentum inside a meson on $J/\psi \to P$ and $\Upsilon \to B_c$ transition form factors and two-body weak hadronic decays of $J/\psi$ and $\Upsilon$ employing the factorization scheme. We predict the branching ratios of semileptonic and nonleptonic weak decays of $J/\psi$ and $\Upsilon$ mesons in Cabibbo-angle-enhanced and Cabibbo-angle-suppressed modes.

# 1 Introduction

Due to remarkable improvements of experimental techniques and instrumentation in the recent years, it is expected that more accurate measurements may now be available for rare decays also. The BES collaboration has observed some rare decays including the semileptonic as well as nonleptonic mode [1, 2]. As a result, it has revived the interest in the rare weak decays of $J/\psi(c\bar{c})$ in to the light quarks, whose branching ratios are expected to be of the order of $10^{-8}$ [3-11]. The future experiments [3, 12-13] of Beijing Electron-Positron Collider (BES-III) and Large Hadron Collider (LHC) hopes to accumulate data for more than $10^{10}$ events of $J/\psi$ per year, which would make it possible to measure such rare decays. From the theoretical point of view, such weak decays are particularly interesting because these are expected to explore mechanism responsible for hadronic transitions, and are also important for the study of nonperturbative QCD effects. Further, decays of a vector meson involve polarization effects that may help in probing the underlying dynamics and hadron structure. Within the Standard Model framework, the flavor changing decays of $J/\psi$ and $\Upsilon$ states are also possible, though naively these are expected to have rather lower branching ratios in comparison to their conventional hadronic and radiative decays.

Earlier, Verma, Kamal and Czarnecki (VKC) [6] had given the first estimates of $J/\psi \to PP/PV$ weak decays using the factorization scheme. VKC employ the Bauer, Stech, and Wirbel (BSW) model to estimate the $J/\psi \to P$ transition form factors and have ignored the $q^2$- dependence in their predictions. Further Sharma and Verma [7] has reanalyzed the $J/\psi$ decays in the same model by including $q^2$- dependence and new values of form factors and decay constants. The predictions in the earlier work [6, 7] are based on $s$-wave dominance for $J/\psi \to PV$. The analysis has also been extended to predict branching ratios of weak decays of $\Upsilon$ based on heavy quark effective theory. Recently, Y. M. Wang *et al.* [8, 9] and Y.L. Shen and Y.M. Wang [10] have calculated $J/\psi \to P$ transition form factors using the QCD sum rules and employing covariant light-front quark model, respectively, to predict the decay rates of $J/\psi$ meson.

In the present work, we employ BSW framework [14] to investigate the effects of flavor dependence on $J/\psi \to P$ and $\Upsilon \to B_c$ transitions form factors and subsiquently on $J/\psi \to PP/PV$ and $\Upsilon \to B_c P/B_c V$ decays, caused by possible variation of average transverse quark momentum ($\omega$) inside a meson. In the light of Heavy Quark Symmetry (HQS) [15], we use the dipole $q^2$-dependence for the form factors $A_0, A_2$ and $V$, and monopole $q^2$-dependence for the form factor $A_1$. We also take into account the contributions from $p-$ and $d-$waves for $J/\psi \to PV$ and $\Upsilon \to B_c P / B_c V$ decays. We predict the branching ratios of weak semileptonic and nonleptonic decays of $J/\psi$ and $\Upsilon$ in Cabibbo-angle-enhanced and Cabibbo-angle-suppressed modes. In support of flavor dependence of form factors and corresponding branching ratios, we also perform an alternate *QCD inspired* calculation to obtain $\omega$ for heavy quarkonium $c\bar{c}\ \&\ b\bar{b}$ states.

The order of presentation is as follows: In Sec. 2 and Sec. 3, we outline the framework employed for analysing semileptonic and nonleptonic weak decays of $J/\psi$. Form factors and branching ratios in BSW model are presented in Sec. 4. The Sec. 5 deals with effects of flavor dependence on $J/\psi$ decays. Semileptonic and nonleptonic weak decays of $\Upsilon$ are analysized in Sec. 6. The major concluson of the present work are discussed and summarized in the last section.

## 2 Semileptonic Weak Decays of $J/\psi$

The semileptonic decay amplitude $A_{SL}(J/\psi \to Pl\bar{\nu})$ can be expressed as

$$A_{SL}(J/\psi \to P) = \frac{G_F}{\sqrt{2}} V_{Qq}^* L^\mu H_\mu \qquad (1)$$

where

$$L^\mu = \nu(k_\nu)\gamma^\mu(1-\gamma_5)l(k_l), \qquad (2)$$

$$H_\mu = \langle P | J_\mu | J/\psi \rangle, \qquad (3)$$

$V_{Qq}$ is the appropriate CKM matrix element for $Q \to q$ transition and $J^\mu$ is the usual weak V-A current. $J/\psi \to P$ matrix element is given by

$$\langle P | J_\mu | J/\psi \rangle = \frac{1}{m_{J/\psi}+m_P} \varepsilon_{\mu\nu\rho\sigma} \varepsilon_{J/\psi}^\nu (P_{J/\psi}+P_P)^\rho q^\sigma V(q^2) - \iota (m_{J/\psi}+m_P) \varepsilon_{J/\psi}^\mu A_1(q^2)$$

$$- \iota \frac{\varepsilon_{J/\psi} \cdot q}{m_{J/\psi}+m_P} (P_{J/\psi}+P_P)^\mu A_2(q^2) + \iota \frac{\varepsilon_{J/\psi} \cdot q}{q^2} (2m_{J/\psi}) q^\mu A_3(q^2)$$

$$-\iota \frac{\varepsilon_{J/\psi} \cdot q}{q^2} (2m_{J/\psi}) q^\mu A_0(q^2), \qquad (4)$$

where $\varepsilon_{J/\psi}$ is the polarization vector of $J/\psi$ and $P_{J/\psi}$ & $P_P$ are the four-momenta of $J/\psi$ and pseudoscalar meson respectively, and $q^\mu = (P_{J/\psi} - P_P)^\mu$. $A_3(q^2)$ is related to $A_1(q^2)$ and $A_2(q^2)$ as

$$A_3(0) = \frac{(m_{J/\psi} + m_P)}{2m_{J/\psi}} A_1(0) + \frac{(m_{J/\psi} - m_P)}{2m_{J/\psi}} A_2(0). \qquad (5)$$

The total decay width for $J/\psi \to P l \bar{\nu}$ is the sum of longitudinal and transverse decay widths given by

$$\Gamma(J/\psi \to P l \bar{\nu}) = \sum_{i=L,\pm} \Gamma_i(J/\psi \to P l \bar{\nu}). \qquad (6)$$

where longitudinal decay width $(\Gamma_L)$ is defined as

$$\Gamma_L(J/\psi \to P l \bar{\nu}) = \frac{G_F^2}{(2\pi)^3} |V_{Qq}|^2 \int_{m_l^2}^{q^2} dq^2 \left(\frac{q^2 - m_l^2}{q^2}\right)^2 \frac{\sqrt{\lambda(m_{J/\psi}^2, m_P^2, q^2)}}{24 m_{J/\psi}^3} \times$$

$$[\frac{3m_l^2}{2q^2} \lambda(m_{J/\psi}^2, m_P^2, q^2) A_0^2(q^2) + \left(1 + \frac{m_l^2}{2q^2}\right) |h_0|^2] \qquad (7a)$$

and the transverse decay width $(\Gamma_T = \Gamma_+ + \Gamma_-)$ is expressed as

$$\Gamma_\pm(J/\psi \to P l \bar{\nu}) = \frac{G_F^2}{(2\pi)^3} |V_{Qq}|^2 \int_{m_l^2}^{q^2} dq^2 \left(\frac{q^2 - m_l^2}{q^2}\right)^2 \frac{\sqrt{\lambda(m_{J/\psi}^2, m_P^2, q^2)}}{24 m_{J/\psi}^3} \left(1 + \frac{m_l^2}{2q^2}\right) |h_\pm|^2. \qquad (7b)$$

The helicity amplitudes $h_0$ and $h_\pm$ are given by

$$h_0 = \frac{(m_{J/\psi}+m_P)}{2m_{J/\psi}}\left\{(m_{J/\psi}^2 - m_P^2 - q^2)A_1(q^2) - \frac{\lambda(m_{J/\psi}^2, m_P^2, q^2)}{(m_{J/\psi}+m_P)^2}A_2(q^2)\right\}, \quad (8)$$

and

$$h_{\pm} = \left\{\frac{\lambda(m_{J/\psi}^2, m_P^2, q^2)}{(m_{J/\psi}+m_P)}V(q^2) \mp (m_{J/\psi}+m_P)A_1(q^2)\right\}, \quad (9)$$

where $m_l$ is the mass of the lepton and $0 \leq q^2 \leq q_{\max}^2 = (m_{J/\psi}-m_P)^2$ and $\lambda(m_{J/\psi}^2, m_P^2, q^2) = (m_{J/\psi}^2 + m_P^2 - q^2) - 4m_{J/\psi}^2 m_P^2$ is related to the three momentum of the daughter meson in the rest frame of $J/\psi$ meson by

$$P_X = \frac{\sqrt{\lambda(m_{J/\psi}^2, m_P^2, q^2)}}{2m_{J/\psi}}. \quad (10)$$

## 3 Nonleptonic Weak Decays of $J/\psi$

### 3.1 Weak Hamiltonian

The QCD modified weak hamiltonian [16] generating the $c-$quark decays in Cabibbo-angle-enhanced mode $(\Delta C = \Delta S = -1)$ is given by

$$H_W = \frac{G_F}{\sqrt{2}}V_{ud}V_{cs}^*\{a_1(\bar{s}c)(\bar{u}d) + a_2(\bar{s}d)(\bar{u}c)\} + h.c., \quad (11a)$$

and for Cabibbo-angle-suppressed mode $(\Delta C = -1, \Delta S = 0)$

$$H_W = \frac{G_F}{\sqrt{2}}V_{ud}V_{cd}^*[a_1\{(\bar{u}d)(\bar{d}c) - (\bar{u}s)(\bar{s}c)\}$$

$$+ a_2\{(\bar{u}c)(\bar{d}d) - (\bar{u}c)(\bar{s}s)\}] + h.c., \quad (11b)$$

where $\bar{q}_1q_2 \equiv \bar{q}_1\gamma_\mu(1-\gamma_5)q_2$ represents the color singlet V-A current and $V_{ij}$ denote standard Cabibbo-Kobayashi-Maskawa (CKM) mixing matrix elements. $a's$ are the undetermined coefficients assigned to the effective charge current, $a_1$, and the effective neutral current, $a_2$, parts of the weak Hamiltonian. These parameters are related to the QCD coefficients $c_{1,2}$ as

$$a_{1,2} = c_{1,2} + \zeta c_{2,1}, \quad (12)$$

where $\zeta = \dfrac{1}{N_c}$, $N_c$ is number of colors. Usually $\zeta$ is treated as a free parameter to be fixed by the experiment. However, we follow the conventional $N_c$ limit to fix the QCD coefficients $a_1 \approx c_1$ and $a_2 \approx c_2$, where [16]

$$a_1 = 1.26\ \&\ a_2 = -0.51, \tag{13}$$

are obtained on the basis of $D \to K\pi$ decays.

## 3.2 Branching Ratios

In the standard factorization scheme, the decay amplitudes of $J/\psi \to PP/PV$ are obtained by sandwiching the QCD modified weak Hamiltonian (upto the weak scale $\dfrac{G_F}{\sqrt{2}} \times CKM factor \times QCD$ coefficient), which are given below as

$$\langle PX | H_W | J/\psi \rangle \sim \langle X | J^\mu | 0 \rangle \langle P | J_\mu^\dagger | J/\psi \rangle, \tag{14}$$

where $X = P, V$. Matrix elements [14] of the weak currents are defined as

$$\langle P(k) | A_\mu | 0 \rangle = -\iota f_P k_\mu, \tag{15a}$$

$$\langle V(k) | V_\mu | 0 \rangle = \varepsilon_\mu^* m_V f_V. \tag{15b}$$

The decay rate formula for $J/\psi \to PP$ decays [17] is given by

$$\Gamma(J/\psi \to PP) = \dfrac{p_c^3}{24\pi m_{J/\psi}^2} | A(J/\psi \to PP) |^2, \tag{16}$$

where $p_c$ is the magnitude of the three momentum of final state meson in the rest frame of $J/\psi$ meson and $m_{J/\psi}$ denote its mass. In general, the three momentum $p_c$ is defined as

$$p_c = \dfrac{1}{2m_{J/\psi}}[m_{J/\psi}^2 - (m_P + m_X)^2][m_{J/\psi}^2 - (m_P - m_X)^2]^{1/2} \tag{17}$$

Combining (8) and (12), the decay amplitude $A(J/\psi \to PP)$, for example, for color-enhanced $J/\psi \to D_{(s)}^+ P$ decays can be expressed as

$$A(J/\psi \to D_{(s)}^+ P) = \dfrac{G_F}{\sqrt{2}} V_{ud} V_{cs}^* a_1 (2m_{J/\psi}) f_\pi A_0^{J/\psi D_{(s)}}(m_P^2). \tag{18}$$

For color-suppressed decays the QCD factor $a_1$ is replaced by $a_2$.

The decay rate [18] of $J/\psi \to PV$ is composed of three independent helicity amplitudes $H_0$ and $H_{\pm 1}$, are expressed as

$$\Gamma(J/\psi \to PV) = \frac{p_c}{8\pi m_{J/\psi}^2}(|H_0|^2 + |H_{+1}|^2 + |H_{-1}|^2). \tag{19}$$

The amplitudes $H_0$ and $H_{\pm 1}$ are defined in terms of the coefficients $a$, $b$ and $c$ as follows:

$$H_{\pm 1} = a \pm c(x^2-1)^{1/2}, \quad H_0 = -ax - b(x^2-1), \tag{20}$$

where

$$x = \frac{m_{J/\psi}^2 - m_P^2 - m_V^2}{2m_{J/\psi}m_V}, \tag{21}$$

$$a = m_V f_V (m_{J/\psi} + m_V) A_1(m_V^2), \tag{22}$$

$$b = -2m_{J/\psi} m_V^2 f_V A_2(m_V^2)/(m_{J/\psi} + m_P), \tag{23}$$

and

$$c = -2m_{J/\psi} m_V^2 f_V V(m_V^2)/(m_{J/\psi} + m_P). \tag{24}$$

The coefficient $a$, $b$ and $c$ describe the $s$-, $d$- and $p$- wave contributions respectively.

## 4  $J/\psi \to P$ Form Factors in BSW Framework

We employ the BSW [14] model for evaluating the meson form factors. In this model, the meson wave function is given by

$$\psi_m(\mathbf{p_T}, x) = N_m \sqrt{x(1-x)} \exp(-\mathbf{p_T}^2/2\omega^2) \exp(-\frac{m^2}{2\omega^2}(x - \frac{1}{2} - \frac{m_{q1}^2 - m_{q2}^2}{2m^2})^2), \tag{25}$$

where $m$ denotes the meson mass and $m_i$ denotes the $i$th quark mass, $N_m$ is the normalization factor and $\omega$ is the average transverse quark momentum, $\omega^2 = \langle \mathbf{p_T}^2 \rangle$. By expressing the current $J_\mu$ in terms of the annihilation and creation operators, the form factors are given by the following integrals:

$$A_0^{J/\psi P}(0) = A_3^{J/\psi P}(0) = \int d^2\mathbf{p_T} \int_0^1 dx (\psi_{J/\psi}^{*1,0}(\mathbf{p_T},x) \sigma_Z^{(0)} \psi_P(\mathbf{p_T},x)), \quad (26)$$

$$V(0) = \frac{m_{q_1(J/\psi)} - m_{q_1(P)}}{m_{J/\psi} - m_P} I, \quad (27)$$

and

$$A_1(0) = \frac{m_{q_1(J/\psi)} + m_{q_1(P)}}{m_{J/\psi} + m_P} I, \quad (28)$$

where

$$I = \sqrt{2} \int d^2\mathbf{p_T} \int_0^1 \frac{dx}{x} (\psi_{J/\psi}^{*1,-1}(\mathbf{p_T},x) i\sigma_y^{(0)} \psi_P(\mathbf{p_T},x)), \quad (29)$$

$m_{q_1(J/\psi)}$ and $m_{q_1(P)}$ denote masses of the non-spectator quarks participating in the quark decay process. From (26) it is clear that the form factors are sensitive to the choice of $\omega$, which is treated as a free parameter. We wish to remark that in BSW [14] model the form factors are calculated by taking same value of $\omega = 0.40$ GeV for initial and final states along with the quark masses (in GeV):

$$m_u = m_d = 0.35, \ m_s = 0.55, \ m_c = 1.7, \text{ and } m_b = 4.9. \quad (30)$$

The $J/\psi \to P$ form factors thus calculated and are presented in rows 2 and 10 of Table 1.

### 4.1 Numerical Results

It has been pointed out in the BSW2 model [15] that consistency with the Heavy Quark Symmetry (HQS) requires certain form factors such as $F_1, A_0, A_2$ and $V$ to have dipole $q^2$ dependence, whereas $A_1$ have monopole $q^2$- dependence i.e.

$$A_0(q^2) = \frac{A_0(0)}{(1 - \frac{q^2}{m_P^2})^2}, \ A_1(q^2) = \frac{A_1(0)}{(1 - \frac{q^2}{m_A^2})},$$

$$A_2(q^2) = \frac{A_2(0)}{(1 - \frac{q^2}{m_A^2})^2}, \text{ and } V(q^2) = \frac{V(0)}{(1 - \frac{q^2}{m_V^2})^2}$$

with appropriate pole masses $m_i$.

### 4.1.1 Branching ratios of semileptonic decays

Using these form factors, we obtain the branching ratios of semileptonic weak decays of $J/\psi \to Pl\bar{\nu}$, and are presented in column 2 of Table 2. The branching ratios $BR_0$, $BR_-$ and $BR_+$ for various $J/\psi \to PV$ corresponding to the contributions of the Helicity amplitudes $h_0$ and $h_{\pm 1}$ are also calculated and are given in columns 2, 3 and 4 of Table 3. We find that

- The branching ratios for semileptonic decays are: $B(J/\psi \to D^+ e^- \bar{V}_e)$ $= (0.23 \times 10^{-8})\%$, $B(J/\psi \to D_s^+ e^- \bar{V}_e) = (3.89 \times 10^{-8})\%$, $B(J/\psi \to D^+ \mu^- \bar{V}_\mu)$ $= (0.22 \times 10^{-8})\%$ and $B(J/\psi \to D_s^+ \mu^- \bar{V}_\mu) = (3.75 \times 10^{-8})\%$

    and are well below the experimental limits.

### 4.1.2 Branching ratios of nonleptonic decays

**(a) $J/\psi \to PP$ Decays**

For $\eta$ and $\eta'$ emitting decays, we take the following basis:

$$\eta = \frac{1}{\sqrt{2}}(u\bar{u} + d\bar{d}) \sin \phi_p - (s\bar{s}) \cos \phi_p, \tag{31a}$$

$$\eta' = \frac{1}{\sqrt{2}}(u\bar{u} + d\bar{d}) \cos \phi_p + (s\bar{s}) \sin \phi_p, \tag{31b}$$

where $\phi_p = \theta_{ideal} - \theta_{physical}$, we take $\theta_{physical} = -15.4°$ [19]. We use the following values for the decay constants (in $GeV$) [19, 20]:

$$f_\pi = 0.131,\ f_K = 0.160,\ f_\eta = 0.133,\ f_{\eta'} = 0.126. \tag{32}$$

Calculated branching ratios for various $J/\psi \to PP$ decays are given in column 2 of Tables 4. It is observed that

- Among the Cabibbo-angle-enhanced $J/\psi \to PP$ decays, the branching ratio of dominant decays are: $B(J/\psi \to D_s^+ \pi^-) = (3.32 \times 10^{-8})\%$, and $B(J/\psi \to D^0 K^0) = (0.72 \times 10^{-8})\%$.

**(b)** $J/\psi \to PV$ **Decays**

We use the following values for the decay constants (in $GeV$) [19, 20]:

$$f_\rho = 0.221, f_{K^*} = 0.220, f_\omega = 0.195, f_\phi = 0.229. \tag{33}$$

Obtained branching ratios for various $J/\psi \to PV$ decays are given in column 2 of Tables 5. For comparison of the relative contributions of the Helicity amplitudes $H_0$ and $H_{\pm 1}$, we have calculated the corresponding branching ratios $BR_0$, $BR_-$ and $BR_+$ for various $J/\psi \to PV$ decays, which are given in column 2 of Table 6. It is observed that

- For the color enhanced decay of the Cabibbo-angle-enhanced mode, the calculated value $B(J/\psi \to D_s^+ \rho^-) = (1.77 \times 10^{-7})\%$, is higher than the branching ratio $B(J/\psi \to D^0 K^{*0}) = (2.51 \times 10^{-8})\%$.

## 5 Flavor dependent effects on $J/\psi$ decays

Since, $\omega$ is a dimensional quantity it may show flavor dependence. Therefore, it may not be justified to take same value of $\omega$ for all the mesons. In our recent work [21], we have investigated the possible flavor dependence through $\omega$ in $B_c \to P/V$ form factors and consequently in $B_c \to PP/PV$ decay widths with improved potential measurement in near future.

### 5.1 Flavor Dependence of $J/\psi \to P$ Form Factors

In this section we investigate the effects of flavor dependence on $J/\psi \to P/V$ decays. Following the prescription of [21], we estimate $\omega$ for different mesons from $|\psi(0)|^2$ i.e. wave function overlap at origin, using the following relation based on the dimensionality arguments

$$|\psi(0)|^2 \propto \omega^3. \tag{34}$$

$|\psi(0)|^2$ is extracted from the hyperfine splitting for the meson masses [22],

$$|\psi(0)|^2 = \frac{9 m_i m_j}{32 \alpha_s \pi}(m_V - m_P), \tag{35}$$

where $m_V$ and $m_P$ respectively denotes masses of vector and pseudoscalar mesons composed of $i$ and $j$ quarks. The meson masses fix quark masses (in GeV) to be $m_u = m_d = 0.31$, $m_s = 0.49$, $m_c = 1.7$, and $m_b = 5.0$ for $\alpha_s(m_b) = 0.19$, $\alpha_s(m_c) = 0.25$, and $\alpha_s = 0.48$ (for light flavors $u$, $d$ and $s$). It may be noted that the uncertainty of $\alpha_s$, particularly for the light quark sector, may lead to variation in quark masses[20,23-25]. In our analysis we allow the following range:

$m_u = m_d = 0.31 \pm 0.05$, $m_s = 0.49 \pm 0.05$, $m_c = 1.7 \pm 0.05$, and $m_b = 5.0 \pm 0.05$.

The calculated numerical values of $|\psi(0)|^2$ are given in column 2 of Table 7. We use the well measured form factor [26] $F_0^{DK}(0) = 0.78 \pm 0.04$ to determine $\omega_D = 0.43 \pm 0.03$ GeV which in turn yields $\omega_{J/\psi} = 0.71 \pm 0.02$ GeV and $\omega$ for other mesons given in column 3 of Table 7. We find that all the $J/\psi \to P$ transition form factors get significantly enhanced due to the flavor dependence of parameter $\omega$. Corresponding uncertainities in form factors due variation in the quark masses are also shown in tables. Note that the uncertainties ranges between 3% to 11% for all the form factors except for $A_2(q^2)$, where the change is roughly 80% to 100%. It may be pointed out that the form factors $A_1(q^2)$, $V(q^2)$ and $A_2(q^2)$ describe the $s$-, $p$- and $d$-wave contributions to the final state vector meson helicity amplitudes, respectively (see eqns (22)-(24)). Thus, $s$-$d$ interference is significant only if both $A_1(q^2)$ and $A_2(q^2)$ are large. At higher $q^2$, amplitudes are dominated by form factor $A_1(q^2)$ (see eqn (20)), and further, if $A_2(q^2)$ is small (as it appears), contributions from the terms propotional to $A_2(q^2)$ are very samll as compared to terms proportional to $A_1(q^2)$. Therefore, the variations even of the order of 80% to 100% are not going to affect the branching ratios by large.

### 5.2 QCD inspired calculation of $\omega$

In this section, we present another method to determine omega (for instance by looking at the typical inverse size of the system under scrutiny, which is of order $\Lambda_{QCD}$ for the lighter mesons and the heavy-light ones and is of order meson mass $(m) \times \alpha_s$ for the heavy quarkonium states. It is interesting to note that following the relation

$$\omega \equiv \alpha_s\, m \text{ for } c\bar{c}\, \&\, b\bar{b},$$

larger values of $\omega$ i.e. $\omega_{J/\psi} = 0.79$ and $\omega_\Upsilon = 1.83$, here we have ignored uncertainities in $\alpha_s$ for heavier quarks. A plot for $\omega$ thus calculated for various mesons (especially for $c\bar{c}\, \&\, b\bar{b}$ states) has interesting pattern w.r.t. the physical masses of the mesons is shown in Fig.1.

It may be pointed out that Sharma and Verma [27] have reportedly shown that $|\psi(0)|^2$ is extracted from leptonic decays in the physical rest mass of meson cannot be simultaneously fitted from $c\bar{c}$ & $b\bar{b}$ in this framework. Perhaps similar analysis needs to be carried out for $|\psi(0)|^2$ extracted from form factors and hyperfine splitting.

### 5.3 Branching ratios including flavor dependent effects

Different values of $\omega$ for initial and final state mesons yield $J/\psi \to P$ transition form factors, which are given in rows 3 and 11 of Table 1. Using these form factors we calculate the branching ratios of semileptonic decays of $J/\psi$ and presented in column 3 of Table 2. Contributions from the Helicity amplitudes $H_0$ and $H_{\pm 1}$ are obtained in the form of branching ratios $BR_0$, $BR_-$ and $BR_+$ for various $J/\psi \to PV$ decays, and are given in column 3 of Table 3. Further, we calculate the branching ratios of various $J/\psi \to PP/PV$ decays, and are listed in column 3 of Tables 4 and 5. In order to compare flavor dependent Helicity amplitudes $H_0$ and $H_{\pm 1}$ with those at $\omega = 0.40$ GeV, we present branching ratios $BR_0$, $BR_-$ and $BR_+$ in column 3 of Table 6. For the sake of comparison, we give results of other works in Tables 2, 4 and 5. The following is observed:

- Branching ratios of all semileptonic as well as nonleptonic decays of $J/\psi$ meson get significantly enhanced because of inclusion of flavor dependent effects.

- The enhanced branching ratios for semileptonic decays are : $B(J/\psi \to D_s^+ e^- \bar{V}_e) = 10.4^{+0.90}_{-0.75} \times 10^{-8}\%$ and $B(J/\psi \to D_s^+ \mu^- \bar{V}_\mu) = 9.93^{+0.95}_{-0.65} \times 10^{-8}\%$. Presently, only experimental [1] upper limits are available,
$$B(J/\psi \to D^+ e^- \bar{V}_e + c.c.) < 1.2 \times 10^{-5},$$
$$B(J/\psi \to D_s^+ e^- \bar{V}_e + c.c.) < 4.9 \times 10^{-5}.$$

- Among the Cabibbo-angle-enhanced $J/\psi \to PP$ decays, we find that branching ratios of dominant decays are: $B(J/\psi \to D_s^+ \pi^-) = 7.41^{+0.13}_{-0.23} \times 10^{-8}\%$, and $B(J/\psi \to D^0 K^0) = 1.39^{+0.10}_{-0.14} \times 10^{-8}\%$.

- For Cabibbo-angle-suppressed but color enhanced modes $J/\psi \to D_s^+ K^-$ and $J/\psi \to D^+ \pi^-$, we obtain the relations based on the naive factorization given as

$$\frac{B(J/\psi \to D_s^+ K^-)}{B(J/\psi \to D_s^+ \pi^-)} = 0.072 \pm 0.003, \text{ and } \frac{B(J/\psi \to D^+ \pi^-)}{B(J/\psi \to D^0 K^0)} = 0.21 \pm 0.03.$$

- In case of $J/\psi \to PV$ decays, for the color enhanced decay of the Cabibbo-angle-enhanced mode, we calculate $B(J/\psi \to D_s^+ \rho^-) = 5.11^{+0.76}_{-0.60} \times 10^{-7}\%$, which is higher than the branching ratio of $J/\psi \to D_s^+ \pi^-$. While for color suppressed decay we obtain $B(J/\psi \to D^0 K^{*0}) = 7.61^{+1.6}_{-1.2} \times 10^{-8}\%$. Our analysis yields

$$\frac{B(J/\psi \to D_s^+ K^{*-})}{B(J/\psi \to D_s^+ \rho^-)} = 0.055 \pm 0.010, \text{ and } \frac{B(J/\psi \to D^+ \rho^-)}{B(J/\psi \to D^0 K^{*0})} = 0.28 \pm 0.07.$$

**5.4 Branching ratios using QCD inspired $\omega$**

Using QCD inspired values of $\omega(\equiv \alpha_s m)$ for heavy quarkonium $c\bar{c}$ & $b\bar{b}$ states, the obtained form factors are given in rows 4 and 12 of Table 1. It is interesting to note all the form factors except $A_0(0)$ get significantly enhanced. Correspondingly, branching ratios of semileptonic and nonleptonic decays are calculated and presented in column 4 of Tables 2 to 6. We observe the following:

- Branching ratios of all semileptonic decays of $J/\psi$ meson get enhanced in comparison to flavor dependent effects. This due to increased value $\omega$ for $c\bar{c}$ state that indicating larger overlap between initial and final state wavefunction.

- Thus, further enhanced branching ratios for semileptonic decays are : $B(J/\psi \to D_s^+ e^- \bar{V}_e) = 10.6 \times 10^{-8}\%$ and $B(J/\psi \to D_s^+ \mu^- \bar{V}_\mu) = 10.2 \times 10^{-8}\%$.

- Branching ratios of all the $J/\psi \to PP$ decays which involve $A_0(0)$ form factor only, are decreased in comparison to flavor dependent branching ratios.

- Branching ratios of all the $J/\psi \to PV$ decays are enhanced in comparison to flavor dependent results.

# 6 Semileptonic and nonleptonic weak decays of $\Upsilon$

## 6.1 $\Upsilon \to B_c$ Form factors and branching ratios

Using the framework described in sections 4 and 5, we obtain the form factor for $\Upsilon \to B_c$ transition at both $\omega = 0.40 \text{GeV}$ and using flavor dependent $\omega$, which are shown in rows 2 and 3 of Table 8. It is observed that flavor dependence significantly enhances the form factors, bringing them close to the expectation [5] based on HQET cosiderations (row 4 of Table 8). Consequently, the branching ratios of semileptonic and nonleptonic weak decays of $\Upsilon$ get significantly enhanced.

We also use the QCD inspired method to determine $\omega$ i.e. $\omega_\Upsilon$=1.83 and subsequently obtain the form factors given in row 4 of Table 8. It may be pointed out that due to marginal change in value of $\omega$ in comparison to flavor dependent value (see Table 7), the change in form factors and branching ratios is negligible, therefore, we exclude these results for further discussion.

## 6.1 Semileptonic weak decays of $\Upsilon$

Using $\Upsilon \to B_c$ form factor, appearing in $b \to c$ transition, and the decay rate formula given in (6), firstly, we calculate the branching ratios for semileptonic decays of $\Upsilon$ at fixed $\omega = 0.40$ GeV and are given in column 2 of Table 9. The predicted branching ratios of semileptonic weak decays of $\Upsilon$ using flavor dependent effects are given in column 3 of Table 9. We find that the branching ratio of dominating semileptonic decay is $B(\Upsilon \to B_c^+ e^- \bar{\nu}_e) = 1.70 \times 10^{-8}\%$ that is sufficiently enhanced in comparison to values at fixed $\omega$ i.e $(6.28 \times 10^{-13}\%)$. Here also we give branching ratios of longitudinal and transverse components ($BR_0$, $BR_-$ and $BR_+$) separately, for semileptonic decays in Table 10.

## 6.2 Nonleptonic weak decays of $\Upsilon$

In this section, the analysis is extended to $\Upsilon \to PP/PV$ decays. The effective weak Hamiltonian generating the dominant b quark decays involving $b \to c$ transition is given by

$$H_W^{\Delta b=1} = \frac{G_F}{\sqrt{2}}\{V_{cb}V_{ud}^*[a_1(\bar{c}b)(\bar{d}u)+a_2(\bar{d}b)(\bar{c}u)]$$

$$+V_{cb}V_{cs}^*[a_1(\bar{c}b)(\bar{s}c)+a_2(\bar{s}b)(\bar{c}c)]\}+h.c., \quad (32)$$

for the CKM favored mode. In our analysis we use $a_1 = 1.12$, $a_2 = -0.26$.

Similar to $J/\psi \to PP$ decays, the factorization scheme expresses weak decay amplitudes as a product of matrix elements of the weak currents (upto the scale $\frac{G_F}{\sqrt{2}} \times (CKM\ factor \times QCD\ factor))$ as:

$$A(\Upsilon \to PP) \sim <P|J^\mu|0><P|J_\mu|\Upsilon>. \quad (33)$$

For instance the decay amplitude for the color enhanced mode $\Upsilon \to B_c^+ \pi^-$ of the CKM-favored decays is given by

$$A(\Upsilon \to B_c^+\pi^-) = \frac{G_F}{\sqrt{2}} V_{cb} V_{ud}^* \, a_1 f_\pi (2m_\Upsilon) A_0^{\Upsilon \to B_c}(m_\pi^2). \quad (34)$$

### 6.1.1 $\Upsilon \to PP$ Decays

Following the same procedure employed in Sections 3 and 4, we calculate the branching ratios for CKM-favored mode both for fixed $\omega = 0.40$ GeV and for flavor dependent $\omega$, and are presented in Table 11 as column 2 and 3 respectively. In addition to the decay constants given in (36) we use $f_D = 0.208$ GeV and $f_{D_s} = 0.273$ GeV. The dominating decays in this mode are found to be: $B(\Upsilon \to B_c^+ D_s^-) = 0.47 \times 10^{-8}\%$ $(2.25 \times 10^{-5}\%)$ and $B(\Upsilon \to B_c^+ \pi^-) = 0.16 \times 10^{-8}\%$ $(7.90 \times 10^{-14}\%)$, where values in the parentheses are branching ratios at fixed $\omega = 0.40$ GeV. It may be noted that our branching ratios including the flavor dependent effects compares well with earlier results obtained by Sharma and Verma [5] (see column 4 of Table 11) using HQET considerations i.e. $B(\Upsilon \to B_c^+ D_s^-) = 0.76 \times 10^{-8}\%$ and $B(\Upsilon \to B_c^+ \pi^-) = 0.33 \times 10^{-8}\%$.

### 6.1.2 $\Upsilon \to PV$ Decays

Employing the decay rate formula for such decays as discussed in the section 3 and following the similar procedure used for $J/\psi \to PV$ decays, we determine the decay amplitudes for various $\Upsilon \to B_c V$ decays for the CKM-favored modes. We use [6] $f_{D^*} = 0.245$ GeV and $f_{D^*_s} = f_{D_s} = 0.273$ GeV to calculate the branching ratios and are given in Table 12. We observe that the dominant mode is $B(\Upsilon \to B_c^+ D_s^{*-}) = 1.79 \times 10^{-8}$ % $(5.01 \times 10^{-13}$ %), following by $B(\Upsilon \to B_c^+ \rho^-) = 0.65 \times 10^{-8}$ % $(2.12 \times 10^{-13}$ %). Here also, the values in the parentheses are branching ratios at $\omega = 0.40$ GeV. For comparison of the contributions of the Helicity amplitudes $H_0$ and $H_{\pm 1}$ involved, we have calculated the corresponding branching ratios $BR_0$, $BR_-$ and $BR_+$ for various $\Upsilon \to PV$ decays both for $\omega = 0.40$ GeV and for flavor dependent $\omega$, which are presented in Table 13.

## 7 Summary and Discussions

In this paper, we have predicted the rare semileptonic and nonleptonic weak decays of $J/\psi$ and $\Upsilon$ meson. It may be mentioned that the present work differs from the previous ones [4, 5] based on the BSW framework in three aspects: Firstly, in the light of HQS based BSW 2 model [12], we use the dipole $q^2$ dependence for the form factors $A_0, A_2$ and $V$, while the earlier work [4, 5] had used the monopole $q^2$ dependence for these form factors. Secondly, the results in the earlier work [5] are based on $s-$ wave dominance, while our results take into account the contributions from $p-$ and $d-$ waves also. Lastly, we incorporate flavor dependent effects on $J/\psi \to P$ and $\Upsilon \to B_c$ form factors in our analysis through the different values of $\omega$ for initial and final state mesons. Also, to support flavor dependence of form factors, we use an alternate QCD inspired approach to determine $\omega$ for heavy quarkonium $c\bar{c}$ & $b\bar{b}$ states. Following conclusions are readily drawn out of our analysis:

A. **Weak decays of $J/\psi$**
- Initially, we calculate $J/\psi \to P$ transition form factors at $\omega = 0.40$ GeV. Among the Cabibbo-angle-enhanced $J/\psi \to PP$ decays, branching ratios of dominant decays are calculated to be $B(J/\psi \to D_s^+ \pi^-) = 3.32 \times 10^{-8}\%$, and $B(J/\psi \to D^0 K^0) = 0.72 \times 10^{-8}\%$. And for Cabibbo-angle-enhanced mode $J/\psi \to PV$ decays, branching ratios of dominant decays are obtained to be $B(J/\psi \to D_s^+ \rho^-) = 1.77 \times 10^{-7}\%$ and $B(J/\psi \to D^0 K^{*-}) = 2.51 \times 10^{-8}\%$.

- We investigate the effects of flavor dependence of $\omega$ by determining $|\psi(0)|^2$ from meson masses to fix $\omega$ for all meson. We, then, calculated $J/\psi \to P$ transition form factors which get significantly enhanced as compared to those at fixed $\omega = 0.40$. As a result of this branching ratios of all semileptonic as well as nonleptonic decays are enhanced.

- Among the Cabibbo-angle-enhanced $J/\psi \to PP/PV$ decays, the dominant decays are predicted as $B(J/\psi \to D_s^+ \pi^-) = 7.41^{+0.13}_{-0.23} \times 10^{-8}\%$, $B(J/\psi \to D^0 K^0) = 1.39^{+0.10}_{-0.14} \times 10^{-8}\%$, $B(J/\psi \to D_s^+ \rho^-) = 5.11^{+0.76}_{-0.60} \times 10^{-7}\%$, and $B(J/\psi \to D^0 K^{*0}) = 7.61^{+1.6}_{-1.2} \times 10^{-8}\%$.

### B. Weak decays of $\Upsilon$

- It is observed that inclusion of flavor dependent effects through $\omega$ factor significantly enhances the form factors and consequently, the branching ratios of semileptonic and nonleptonic weak decays of $\Upsilon$.

- Branching ratio of the dominating semileptonic mode is predicted to be $B(\Upsilon \to B_c^+ e^- \bar{\nu}_e) = 1.70 \times 10^{-8}\%$.

- Among the Cabibbo-angle-enhanced $\Upsilon \to B_c P / B_c V$ decay modes, the dominant one are predicted as $B(\Upsilon \to B_c^+ D_s^-) = 0.47 \times 10^{-8}\%$, $B(\Upsilon \to B_c^+ \pi^-) = 0.16 \times 10^{-8}\%$, $B(\Upsilon \to B_c^+ D_s^{*-}) = 1.79 \times 10^{-8}\%$ and $B(\Upsilon \to B_c^+ \rho^-) = 0.65 \times 10^{-8}\%$.

- Including the flavor dependent effects through $\omega$ and this prescription, it is found that the our predictions agree well with those obtained by Sharma and Verma [5] using the HQET considerations.

- Though the predicted branching ratios in QCD inspired method of obtaining $\omega$ are marginally changed, the alternate calculation of $\omega$ for heavy quarkonium $c\bar{c}\ \&\ b\bar{b}$ states seems to support the flavor dependent effects on rare weak decays of $J/\psi$ and $\Upsilon$ mesons.

It is hoped that these branching ratios would lie in the detectable range and may be measured in future experiments.

**Acknowledgements**

Financial assistance from UGC, New Delhi is gratefully acknowledged.

**References**

1. BES Collaboration, M. Ablikim *et al.*, Phys. Lett. B**639**, 418 (2006).

2. BES Collaboration, M. Ablikim *et al.*, Phys. Lett. B**663**, 297 (2008)

3. Hai-Bo Li and SHi-Hai Zhu, arXiv:1202.2955v2 [hep-ex] (2012) and references therein.

4. Hong-Wei Ke, Ya-Zheng Chen, and Xue-Qian Li, Chin. Phys. Lett. **28**, 071301 (2011).

5. M.A. Sanchis-Lonzano, Z. Phys. C **62**, 271 (1994).

6. R.C. Verma, A.N. Kamal, and A. Czarnecki, Phys. Lett. B**252**, 690 (1990).

7. K.K. Sharma and R.C. Verma, Int. J. Mod. Phys. A **14**, 937 (1999).

8. Y.M. Wang, H. Zou, Z.T. Wei, X.Q. Li, and C.D. Lu, Eur. Phys. J. C **54**, 107 (2008).

9. Y.M. Wang, H. Zou, Z.T. Wei, X. Q. Li, and C.D. Lu, Eur. Phys. J. C **55**, 607 (2008).

10. Y.L. Shen and Y.M. Wang, Phys. Rev. D **78**, 074012 (2008).

11. Y. M. Wang, H. Zou, Z. T. Wei, X. Q. Li and C. D. Lu, J. Phys. G **36**, 105002 (2009).

12. D.M. Asner *et al.*, Report No. IHEP-Physics-Report-BES-III-2008-001, arXiv:0809.1869v1.

13. G.T. Bodwin *et al.*, CERN YELLOW REPORT, CERN-2005-005, Geneva 487(2005), [hep-ex/0412158]; I.C. Arsene *et al.*, Phys. Lett. B**660**, 176 (2008).

14. M. Wirbel, B. Stech and M. Bauer, Z. Phys. C **29**, 637 (1985); M. Bauer, B. Stech and M. Wirbel, Z. Phys. C **34**, 103 (1987); M. Wirbel, Prog. Part. Nucl. Phys. **21**, 33 (1988).

15. M. Neubert et al., "Exclusive weak decays of B-mesons", ed. A.J. Buras and H.

Linder (Singapore: World Scientific), **28** (1992); H.Y. Cheng, Phys. Rev. D **67**, 094007 (2003); Phys. Rev. D **69**, 074025 (2004) and references therein.

16. T.E. Browder and K. Honscheid, Prog. Part. Nucl. Phys. **35**, 81 (1995); M. Neubert, V. Rieckert, B. Stech and Q.P. Xu, in Heavy Flavours, edited by A. J. Buras and H. Lindner, World Scientific, Singapore, 1992, and references therein.

17. A. Ali, J.G. Korner, G. Kramer and J. Willordt, Z. Phys. C **1**, 269 (1979); A. Ali, J.G. Korner, G. Kramer and J. Willordt, Z. Phys. C **2**, 33 (1979)

18. G. Kramer and W.F. Palmer, Phys. Rev. D **45**, 193 (1992); X. Li, G. Lu, and Y.D. Yang, Phys. Rev. D **68**, 114015 (2003).

19. Particle Data Group: C. Amsler *et al.*, Phys. Lett. B**667**, 1 (2008).

20. H.Y. Cheng, Phys. Rev. D **67** (2003) 094007; Phys. Rev. D **69** (2004) 074025, and references therein.

21. R. Dhir, N. Sharma and R.C. Verma, J. Phys. G. **35**, 085002 (2008); R. Dhir and R.C. Verma, Phys. Rev. D **79**, 034004 (2009).

22. D.H. Perkins, "Introduction to High Energy Physics", 4th edition, Cambridge University Press (2000).

23. H.Y. Cheng *et al*, *Phys. Rev.* D **55**, 1559 (1997) and references therein.

24. W. Jaus, Phys. Rev. D **53**, 1349 (1996); *ibid.* **54**, 5904 (1996).

25. M. Karliner and H.J. Lipkin, Report: TAUP 2735–03 (2003) (*Preprint* hep-ph/0307243v); M. Karliner and H.J. Lipkin, Report:TAUP 2829/06, WIS/07/06-JULY-DPP, ANL-HEP-PR-06–56 (2007), *Preprint* hep-ph/0608004v3, and references therein.

26. M. Ablikim *et al.*, (BES), Phys. Lett. B**597**, 39 (2004); [hep-ex/0406028]; D.Y. Kim, Nucl. Phys. B**167**, 75 (2007); [hep-ex/0609046]; R. Dhir and R.C. Verma, J. Phys. G. **34**, 637 (2007); and references therein.

27. A.C. Sharma, R.C. Verma and M.P. Khanna, Ind. J. Pure & App. Phys. **36**, 259 (1998).

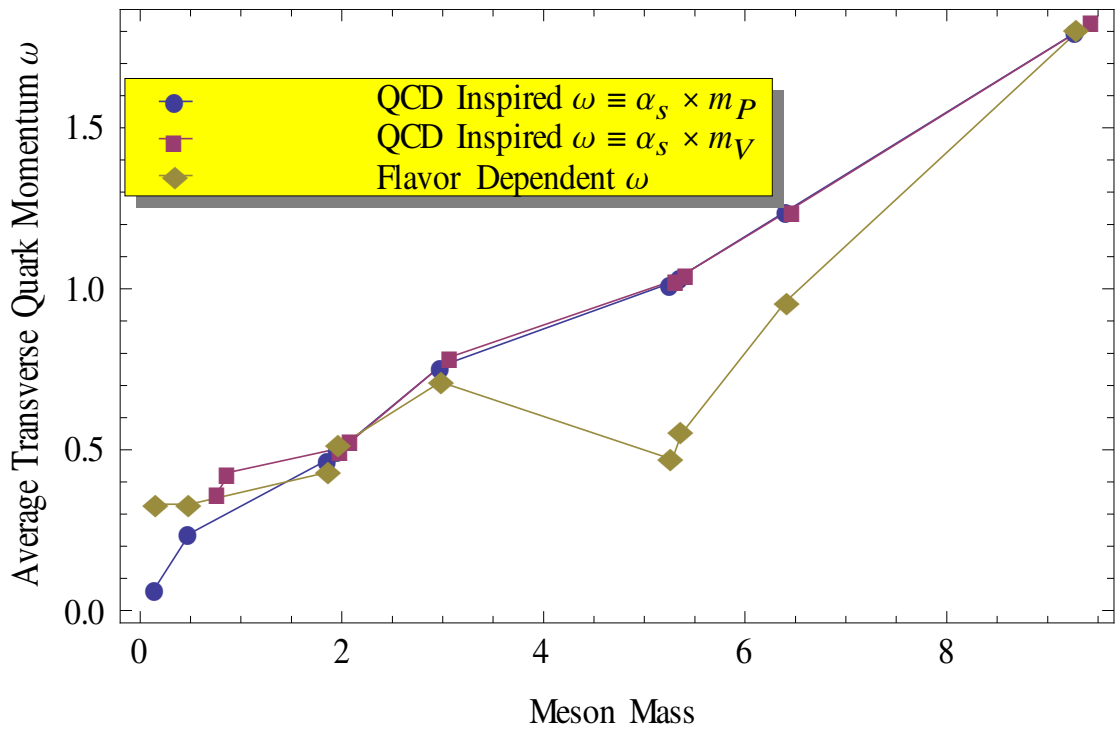

**Figure 1.** Variation of average transeverse quark momentum w.r.t. physical meson masses

Table 1: Form factors of $J/\psi \to P$ transitions

| | Models | $A_0^{J/\psi D}(0)$ | $A_1^{J/\psi D}(0)$ | $A_2^{J/\psi D}(0)$ | $V^{J/\psi D}(0)$ |
|---|---|---|---|---|---|
| This Work | $\omega = 0.40$ GeV | 0.40 | 0.44 | -0.23 | 1.17 |
| | Using flavor dependent $\omega$ | 0.55±0.02 | $0.77^{+0.09}_{-0.07}$ | $0.31^{+0.27}_{-0.16}$ | $2.14^{+0.15}_{-0.11}$ |
| | QCD inspired $\omega \equiv \alpha_s m$ | 0.54 | 0.80 | 0.47 | 2.21 |
| | [5] | 0.61 | 0.68 | -0.28 | 1.82 |
| | [7] | 0.27 | 0.27 | 0.34 | 0.81 |
| | [8] | 0.68 | 0.68 | 0.18 | 1.6 |

| | Models | $A_0^{J/\psi D_s}(0)$ | $A_1^{J/\psi D_s}(0)$ | $A_2^{J/\psi D_s}(0)$ | $V^{J/\psi D_s}(0)$ |
|---|---|---|---|---|---|
| This Work | $\omega = 0.40$ GeV | 0.47 | 0.55 | -0.14 | 1.26 |
| | Using flavor dependent $\omega$ | $0.71^{+0.04}_{-0.02}$ | 0.94±0.07 | $0.33^{+0.33}_{-0.23}$ | $2.30^{+0.09}_{-0.06}$ |
| | QCD inspired $\omega \equiv \alpha_s m$ | 0.69 | 0.96 | 0.51 | 2.36 |
| | [5] | 0.66 | 0.78 | -0.10 | 1.80 |
| | [7] | 0.37 | 0.38 | 0.35 | 1.07 |
| | [8] | 0.68 | 0.68 | 0.13 | 1.8 |

Table 2: Branching Ratios (in the units of $10^{-10}$) of $J/\psi \to Pl\bar{\nu}$ decays

| | This Work | | | | |
|---|---|---|---|---|---|
| Decays | $\omega = 0.40$ GeV | Using Flavor Dependent $\omega$ | QCD inspired $\omega \equiv \alpha_s m$ | [7] | [8] |
| $J/\psi \to D^+ e^- \bar{\nu}_e$ | 0.23 | $0.60^{+0.08}_{-0.07}$ | 0.62 | 0.073 | 5.4 |
| $J/\psi \to D_s^+ e^- \bar{\nu}_e$ | 3.89 | $10.4^{+0.90}_{-0.75}$ | 10.6 | 1.8 | 5.6 |
| $J/\psi \to D^+ \mu^- \bar{\nu}_\mu$ | 0.22 | $0.58^{+0.08}_{-0.06}$ | 0.60 | 0.071 | 5.1 |
| $J/\psi \to D_s^+ \mu^- \bar{\nu}_\mu$ | 3.75 | $9.93^{+0.95}_{-0.65}$ | 10.2 | 1.7 | 5.6 |

Table 3: Branching Ratios $BR_0$, $BR_-$ and $BR_+$ corresponding to the Helicity amplitudes $h_0$ and $h_{\pm 1}$ (in the units of $10^{-10}$) of $J/\psi \to Pl\bar{\nu}$ decays

| Decays | $\omega = 0.40$ GeV | | | Using flavor dependent $\omega$ | | | QCD inspired $\omega \equiv \alpha_s\, m$ | | |
|---|---|---|---|---|---|---|---|---|---|
| | $BR_0$ | $BR_-$ | $BR_+$ | $BR_0$ | $BR_-$ | $BR_+$ | $BR_0$ | $BR_-$ | $BR_+$ |
| $J/\psi \to D^+ e^- \bar{\nu}_e$ | 0.14 | 0.076 | 0.014 | $0.31 \pm 0.04$ | $0.24^{+0.04}_{-0.03}$ | $0.048^{+0.005}_{-0.004}$ | 0.31 | 0.26 | 0.052 |
| $J/\psi \to D_s^+ e^- \bar{\nu}_e$ | 2.65 | 0.11 | 0.19 | $6.42 \pm 0.40$ | $3.25^{+0.42}_{-0.20}$ | $0.59^{+0.06}_{-0.05}$ | 6.40 | 3.54 | 0.63 |
| $J/\psi \to D^+ \mu^- \bar{\nu}_\mu$ | 0.14 | 0.074 | 0.013 | $0.30 \pm 0.04$ | $0.24^{+0.04}_{-0.03}$ | $0.047 \pm 0.005$ | 0.30 | 0.36 | 0.050 |
| $J/\psi \to D_s^+ \mu^- \bar{\nu}_\mu$ | 2.55 | 0.10 | 0.19 | $6.17 \pm 0.41$ | $3.26^{+0.33}_{-0.25}$ | $0.58^{+0.07}_{-0.05}$ | 6.14 | 3.44 | 0.61 |

Table 4: Branching Ratios (in the units of $10^{-10}$) of $J/\psi \to PP$ decays

| Decay | This Work | | | [5] | [7] | [8] |
|---|---|---|---|---|---|---|
| | $\omega = 0.40$ GeV | Using flavor dependent $\omega$ | QCD inspired $\omega \equiv \alpha_s\, m$ | | | |
| $\Delta C = \Delta S = +1$ | | | | | | |
| $J/\psi \to D_s^+ \pi^-$ | 3.32 | $7.41^{+0.13}_{-0.23}$ | 7.13 | 6.12 | 2.0 | 2.5 |
| $J/\psi \to D^0 K^0$ | 0.72 | $1.39^{+0.10}_{-0.14}$ | 1.34 | 1.96 | 0.36 | 0.5 |
| $\Delta C = +1, \Delta S = 0$ | | | | | | |
| $J/\psi \to D_s^+ K^-$ | 0.24 | $0.53 \pm 0.02$ | 0.52 | 0.39 | 0.16 | – |
| $J/\psi \to D^+ \pi^-$ | 0.15 | $0.29^{+0.02}_{-0.03}$ | 0.028 | 0.39 | 0.08 | – |
| $J/\psi \to D^0 \pi^0$ | 0.012 | $0.024^{+0.002}_{-0.003}$ | 0.023 | 0.039 | – | – |
| $J/\psi \to D^0 \eta$ | 0.036 | $0.070^{+0.005}_{-0.007}$ | 0.067 | 0.011 | – | – |
| $J/\psi \to D^0 \eta'$ | 0.002 | $0.004^{+0.0002}_{-0.0005}$ | 0.004 | 0.002 | – | – |
| $\Delta C = +1, \Delta S = -1$ | | | | | | |
| $J/\psi \to D^+ K^-$ | 0.012 | $0.023 \pm 0.002$ | 0.022 | – | – | – |
| $J/\psi \to D^0 \bar{K}^0$ | 0.002 | $0.004^{+0.000}_{-0.001}$ | 0.004 | – | – | – |

Table 5: Branching Ratios (in the units of $10^{-10}$) of $J/\psi \to PV$ decays

| Decay | This Work $\omega = 0.40$ GeV | This Work Using flavor dependent $\omega$ | This Work QCD inspired $\omega \equiv \alpha_s m$ | [5] | [7] | [8] |
|---|---|---|---|---|---|---|
| $\Delta C = \Delta S = +1$ | | | | | | |
| $J/\psi \to D_s^+ \rho^-$ | 17.7 | $51.1^{+7.6}_{-6.0}$ | 53.2 | 25.41 | 12.6 | 28 |
| $J/\psi \to D^0 K^{*0}$ | 2.51 | $7.61^{+1.6}_{-1.2}$ | 8.12 | 7.19 | 1.54 | 5.5 |
| $\Delta C = +1, \Delta S = 0$ | | | | | | |
| $J/\psi \to D_s^+ K^{*-}$ | 0.97 | $2.82^{+0.40}_{-0.30}$ | 2.96 | 1.48 | 0.82 | - |
| $J/\psi \to D^+ \rho^-$ | 0.72 | $2.16^{+0.5}_{-0.3}$ | 2.28 | 1.54 | 0.42 | - |
| $J/\psi \to D^0 \rho^0$ | 0.06 | $0.18 \pm 0.03$ | 0.19 | 0.15 | - | - |
| $J/\psi \to D^0 \omega$ | 0.05 | $0.16^{+0.03}_{-0.02}$ | 0.17 | 0.13 | - | - |
| $J/\psi \to D^0 \phi$ | 0.14 | $0.42^{+0.08}_{-0.07}$ | 0.44 | 0.46 | - | - |
| $\Delta C = +1, \Delta S = -1$ | | | | | | |
| $J/\psi \to D^+ K^{*-}$ | 0.042 | $0.13^{+0.02}_{-0.03}$ | 0.13 | – | – | – |
| $J/\psi \to D^0 \overline{K}^{*0}$ | 0.007 | $0.021^{+0.004}_{-0.003}$ | 0.022 | – | – | – |

Table 6: Branching Ratios $BR_0$, $BR_-$ and $BR_+$ corresponding to the Helicity amplitudes $H_0$ and $H_{\pm 1}$ (in the units of $10^{-10}$) of $J/\psi \to PV$ decays

| Decay | $\omega = 0.40$ GeV | | | Using flavor dependent $\omega$ | | | QCD inspired $\omega \equiv \alpha_s m$ | | |
|---|---|---|---|---|---|---|---|---|---|
| | $BR_0$ | $BR_+$ | $BR_-$ | $BR_0$ | $BR_+$ | $BR_-$ | $BR_0$ | $BR_+$ | $BR_-$ |
| $\Delta C = \Delta S = +1$ | | | | | | | | | |
| $J/\psi \to D_s^+ \rho^-$ | 6.26 | 2.75 | 8.74 | $17.3^{+2.3}_{-1.8}$ | $7.55^{+0}_{-0.30}$ | $26.3^{+3.6}_{-2.7}$ | 17.7 | 7.97 | 27.7 |
| $J/\psi \to D^0 K^{*0}$ | 0.75 | 0.32 | 1.45 | $2.13^{+0.42}_{-0.35}$ | $0.93^{+0.18}_{-0.20}$ | $4.56^{+0.86}_{-0.67}$ | 2.25 | 1.00 | 4.85 |
| $\Delta C = +1, \Delta S = 0$ | | | | | | | | | |
| $J/\psi \to D_s^+ K^{*-}$ | 0.27 | 0.18 | 0.52 | $0.74^{+0.10}_{-0.08}$ | $0.52^{+0.10}_{-0.08}$ | $1.56^{+0.42}_{-0.35}$ | 0.77 | 0.54 | 1.65 |
| $J/\psi \to D^+ \rho^-$ | 0.027 | 0.076 | 0.37 | $0.77^{+0.15}_{-0.12}$ | $0.22^{+0.06}_{-0.05}$ | $1.17^{+0.22}_{-0.17}$ | 0.81 | 0.24 | 1.25 |
| $J/\psi \to D^0 \rho^0$ | 0.022 | 0.066 | 0.031 | $0.064^{+0.012}_{-0.011}$ | $0.018^{+0.005}_{-0.004}$ | $0.096^{+0.018}_{-0.014}$ | 0.067 | 0.019 | 0.103 |
| $J/\psi \to D^0 \omega$ | 0.020 | 0.006 | 0.028 | $0.056^{+0.011}_{-0.019}$ | $0.017^{+0.005}_{-0.004}$ | $0.088^{+0.022}_{-0.013}$ | 0.059 | 0.018 | 0.094 |
| $J/\psi \to D^0 \phi$ | 0.032 | 0.022 | 0.083 | $0.091^{+0.018}_{-0.015}$ | $0.063^{+0.018}_{-0.013}$ | $0.26^{+0.05}_{-0.04}$ | 0.096 | 0.068 | 0.28 |
| $\Delta C = +1, \Delta S = -1$ | | | | | | | | | |
| $J/\psi \to D^+ K^{*-}$ | 0.012 | 0.0053 | 0.024 | $0.035^{+0.007}_{-0.006}$ | $0.015^{+0.005}_{-0.003}$ | $0.075^{+0.014}_{-0.011}$ | 0.037 | 0.016 | 0.080 |
| $J/\psi \to D^0 \overline{K}^{*0}$ | 0.002 | 0.004 | 9 | $0.006^{+0.001}_{-0.001}$ | $0.003^{+0.000}_{-0.001}$ | $0.012^{+0.003}_{-0.001}$ | 0.006 | 0.003 | 0.013 |

Table 7: $|\psi(0)|^2$ and $\omega$ for vector and pseudoscalar mesons

| Meson | $\|\psi(0)\|^2$ (in GeV$^3$) | Parameter $\omega$ (in GeV) | QCD inspired $\omega \equiv \alpha_s\, m$ (in GeV) |
|---|---|---|---|
| $\rho(\pi)$ | $0.011 \pm 0.003$ | $0.33 \pm 0.03$ | - |
| $K^*(K)$ | $0.011 \pm 0.03$ | $0.33 \pm 0.03$ | - |
| $D^*(D)$ | $0.026^{+0.006}_{-0.004}$ | $0.43 \pm 0.03$ | - |
| $D_s^*(D_s)$ | $0.041^{+0.007}_{-0.005}$ | $0.51^{+0.02}_{-0.03}$ | - |
| $J/\psi(\eta_c)$ | $0.115^{+0.011}_{-0.008}$ | $0.71 \pm 0.02$ | 0.79 |
| $B^*(B)$ | $0.033 \pm 0.005$ | $0.47^{+0.02}_{-0.03}$ | - |
| $B_s^*(B_s)$ | $0.053 \pm 0.006$ | $0.55 \pm 0.02$ | - |
| $B_c(B_c^*)$ | $0.281^{+0.0017}_{-0.0011}$ | $0.96 \pm 0.02$ | - |
| $\Upsilon(\eta_b)$ | $1.850^{+0.03}_{-0.04}$ | $1.80^{+0.01}_{-0.02}$ | 1.83 |

Table 8: Form factors of $\Upsilon \to B_c$ transitions

| | Models | $A_0^{\Upsilon B_c}(0)$ | $A_1^{\Upsilon B_c}(0)$ | $A_2^{\Upsilon B_c}(0)$ | $V^{\Upsilon B_c}(0)$ |
|---|---|---|---|---|---|
| This Work | $\omega = 0.40$ GeV | 0.003 | 0.003 | 0.003 | 0.008 |
| | Using flavor dependent $\omega$ | $0.46 \pm 0.01$ | $0.62^{+0.02}_{-0.01}$ | $0.38^{+0.06}_{-0.09}$ | $1.61 \pm 0.01$ |
| | QCD inspired $\omega \equiv \alpha_s\, m$ * | 0.46 | 0.62 | 0.38 | 1.62 |
| | [5] | 0.98 | 1.01 | 1.01 | 1.01 |

*the form factors here have shown no significant change in the values therefor the BRs for the decays involving $\Upsilon \to B_c$ form factors for **QCD inspired** $\omega$ remains same as for flavor dependent $\omega$

Table 9: Branching Ratios (in the units of $10^{-10}$) of $\Upsilon \to Pl\bar{\nu}$ decays

| Decays | This Work | |
| --- | --- | --- |
| | $\omega = 0.40$ GeV | Using Flavor Dependent $\omega$ |
| $\Upsilon \to B_c^+ e^- \bar{\nu}_e$ | $6.28 \times 10^{-5}$ | $1.70^{+0.03}_{-0.02}$ |
| $\Upsilon \to B_c^+ \mu^- \bar{\nu}_\mu$ | $6.24 \times 10^{-5}$ | $1.69^{+0.04}_{-0.02}$ |
| $\Upsilon \to B_c^+ \tau^- \bar{\nu}_\tau$ | $9.86 \times 10^{-5}$ | $0.29^{+0.05}_{-0.02}$ |

Table 10: Branching Ratios $BR_0$, $BR_-$ and $BR_+$ corresponding to the Helicity amplitudes $h_0$ and $h_{\pm 1}$ (in the units of $10^{-10}$) of $\Upsilon \to Pl\bar{\nu}$ decays

| Decays | $\omega = 0.40$ GeV | | | Using flavor dependent $\omega$ | | |
| --- | --- | --- | --- | --- | --- | --- |
| | $BR_0$ | $BR_-$ | $BR_+$ | $BR_0$ | $BR_-$ | $BR_+$ |
| $\Upsilon \to B_c^+ e^- \bar{\nu}_e$ | $5.50 \times 10^{-5}$ | $0.46 \times 10^{-5}$ | $3.10 \times 10^{-5}$ | $1.40^{+0.03}_{-0.02}$ | $0.18^{+0.01}_{-0.00}$ | $0.13^{+0.01}_{-0.00}$ |
| $\Upsilon \to B_c^+ \mu^- \bar{\nu}_\mu$ | $5.47 \times 10^{-5}$ | $0.45 \times 10^{-5}$ | $3.08 \times 10^{-5}$ | $1.39^{+0.03}_{-0.02}$ | $0.18^{+0.01}_{-0.00}$ | $0.12^{+0.01}_{-0.00}$ |
| $\Upsilon \to B_c^+ \tau^- \bar{\nu}_\tau$ | $0.86 \times 10^{-5}$ | $0.81 \times 10^{-6}$ | $0.45 \times 10^{-6}$ | $0.24^{+0.00}_{-0.01}$ | $0.03$ | $0.018$ |

Table 11: Branching Ratios (in the units of $10^{-10}$) of $\Upsilon \to B_c P$ decays

| Decays | This Work | | [5] |
| --- | --- | --- | --- |
| | $\omega = 0.40$ GeV | Using flavor dependent $\omega$ | |
| $\Delta b = 1, \Delta C = 1, \Delta S = 0$ | | | |
| $\Upsilon \to B_c^+ \pi^-$ | $7.90 \times 10^{-6}$ | $0.16 \pm 0.01$ | $0.33$ |
| $\Delta b = 1, \Delta C = 0, \Delta S = -1$ | | | |
| $\Upsilon \to B_c^+ D_s^-$ | $2.25 \times 10^{-5}$ | $0.47 \pm 0.01$ | $0.76$ |
| $\Delta b = 1, \Delta C = 1, \Delta S = -1$ | | | |
| $\Upsilon \to B_c^+ K^-$ | $6.06 \times 10^{-7}$ | $0.013$ | $0.024$ |
| $\Delta b = 1, \Delta C = 0, \Delta S = 0$ | | | |
| $\Upsilon \to B_c^+ D^-$ | $7.25 \times 10^{-7}$ | $0.015^{+0.01}_{-0.00}$ | $0.031$ |

Table 12: Branching Ratios (in the units of $10^{-10}$) of $\Upsilon \to B_c V$ decays

| Decays | This Work | | [5] |
|---|---|---|---|
| | $\omega = 0.40$ GeV | Using flavor dependent $\omega$ | |
| $\Delta b = 1, \Delta C = 1, \Delta S = 0$ | | | |
| $\Upsilon \to B_c^+ \rho^-$ | $2.12 \times 10^{-5}$ | $0.65 \pm 0.03$ | 0.88 |
| $\Delta b = 1, \Delta C = 0, \Delta S = -1$ | | | |
| $\Upsilon \to B_c^+ D_s^{*-}$ | $5.01 \times 10^{-5}$ | $1.79^{+0.09}_{-0.04}$ | 2.57 |
| $\Delta b = 1, \Delta C = 1, \Delta S = -1$ | | | |
| $\Upsilon \to B_c^+ K^{*-}$ | $1.14 \times 10^{-6}$ | $0.035 \pm 0.02$ | 0.050 |
| $\Delta b = 1, \Delta C = 0, \Delta S = 0$ | | | |
| $\Upsilon \to B_c^+ D^{*-}$ | $2.06 \times 10^{-6}$ | $0.073^{+0.003}_{-0.002}$ | 0.11 |

Table 13: Branching Ratios $BR_0$, $BR_-$ and $BR_+$ corresponding to the Helicity amplitudes $H_0$ and $H_{\pm 1}$ (in the units of $10^{-10}$) of $\Upsilon \to B_c V$ decays

| Decays | $\omega = 0.40$ GeV | | | Using flavor dependent $\omega$ | | |
|---|---|---|---|---|---|---|
| | $BR_0$ | $BR_-$ | $BR_+$ | $BR_0$ | $BR_-$ | $BR_+$ |
| $\Delta b = 1, \Delta C = 1, \Delta S = 0$ | | | | | | |
| $\Upsilon \to B_c^+ \rho^-$ | $1.82 \times 10^{-5}$ | $2.48 \times 10^{-6}$ | $6.53 \times 10^{-7}$ | $0.53^{+0.02}_{-0.01}$ | $0.09^{+0.01}_{-0.00}$ | $0.024^{+0.02}_{-0.01}$ |
| $\Delta b = 1, \Delta C = 0, \Delta S = -1$ | | | | | | |
| $\Upsilon \to B_c^+ D_s^{*-}$ | $1.94 \times 10^{-5}$ | $2.30 \times 10^{-5}$ | $7.73 \times 10^{-6}$ | $0.62^{+0.02}_{-0.01}$ | $0.88^{+0.04}_{-0.02}$ | $0.29^{+0.01}_{-0.01}$ |
| $\Delta b = 1, \Delta C = 1, \Delta S = -1$ | | | | | | |
| $\Upsilon \to B_c^+ K^{*-}$ | $9.20 \times 10^{-5}$ | $1.71 \times 10^{-8}$ | $4.55 \times 10^{-8}$ | $0.027^{+0.001}_{-0.000}$ | $0.007^{+0.001}_{-0.001}$ | 0.017 |
| $\Delta b = 1, \Delta C = 0, \Delta S = 0$ | | | | | | |
| $\Upsilon \to B_c^+ D^{*-}$ | $8.75 \times 10^{-7}$ | $2.28 \times 10^{-7}$ | $2.97 \times 10^{-7}$ | $0.027^{+0.001}_{-0.001}$ | $0.035^{+0.001}_{-0.001}$ | $0.010^{+0.002}_{-0.001}$ |